\documentclass[10pt,letterpaper]{article}
\usepackage{opex3}
\usepackage{amsmath}
\usepackage{tabularx}
\usepackage{upgreek}

\begin{document}

\title{Analytical description of interference between two misaligned and mismatched complete Gaussian beams}

\author{Gudrun Wanner$^{*}$,  Gerhard Heinzel}

\address{Max Planck Institute for Gravitational Physics (Albert Einstein Institute) and 
Institute for Gravitational Physics of the Leibniz Universit\"at Hannover, Callinstra§e 38, D-30167 Hannover, Germany}

\email{*Gudrun Wanner: gudrun.wanner@aei.mpg.de} 


\begin{abstract}
A typical application for laser interferometers is a precision measurement of length changes that result in interferometric phase shifts. Such phase changes are typically predicted numerically, due to the complexity of the overlap integral that needs to be solved.  
In this paper we will derive analytical representations of the interferometric phase and contrast (aka. fringe visibility) for two beam interferometers, both homodyne and heterodyne. The fundamental Gaussian beams can be arbitrarily misaligned and mismatched to each other. A limitation of the analytical result is that both beams must be detected completely, which can experimentally be realized by a sufficiently large single-element photodetector.
\end{abstract}



\section{Introduction}
Phase shifts in laser interferometers are a precision measure for length variations. These phase shifts are typically predicted using either commercial software tools such as ZEMAX, CodeV, FRED and alike, which often compute phases with respect to either a reference sphere or a reference plane. Alternatively dedicated algorithms are used in academic environments to compute the interferometer phase (see for example \cite{Wanner2012, Andreas2012}). In either case, a variety of different methods are applied, such as Fourier optics, or Gaussian beam propagation with the ABCD formalism followed by numerical integration and final phase computation by fringe analyses methods \cite{Cassaing2001, Surrel2000, Surrel1997, Surrel1993, Bruning1974, Larkin1992, Freischlad1990}.
 In any chosen option, the phase shift is computed numerically. We will show here, that it is also possible to compute the interferometric phase and contrast on a large single element photodiode analytically, for two fundamental Gaussian beams with arbitrary beam parameters (waist positions and waist sizes) and arbitrary mutual alignment. 
The only assumptions made here are:
\begin{enumerate}  
	\item The detector is an infinite plane; in our experience this a valid assumption if the detector area covers at least three times the Gaussian radius of both beams such that no clipping occurs.
	\item On the detector the spot sizes $w_{m,r}$, radii of curvature $R_{m,r}$ and Gouy phases $\eta_{m,r}$ are unaffected by applied shifts and tilts. These parameters remain constant during coordinate transformations. This is a valid assumption for the usual case of small misalignments.
	\item  Both beams have the same wave number $k_m=k_r=: k$ which is the case if the interferometer is either 
	\begin{itemize}
		 \item homodyne (the interfering beams have identical (angular) laser frequencies $\omega_m \equiv \omega_r$) or 
		\item  heterodyne with a heterodyne frequency $\Delta \omega := |\omega_m - \omega_r|$ which is small compared to the laser frequencies: $\Delta \omega \ll \omega_{m,r}$\;.
	\end{itemize}	
\end{enumerate}
The first assumption is the most restrictive, valid only when both beams are completely sensed by the detector. It is violated, e.g., by quadrant photodiodes due to their insensitive slits separating the segments, and the equations derived here do not apply.
%
\section{Fundamentals}\label{sec:Fundamental}
The electric field of a laser beam in fundamental Gaussian mode can be described using a phase $\Phi$ and a real valued beam amplitude $A$: 
\begin{equation}
	E(r_b,z_b,t)= A(r_b,z_b) \exp(i \omega t - i \Phi(r_b,z_b))\;,
\end{equation}
with the local beam coordinates $r_b,z_b$ and
\begin{align}
	A(r_b,z_b)	&=\sqrt{2 Z} \sqrt{\frac{2 P}{\pi w^2(z_b)}}\exp\left(\frac{-r_b^2}{w^2(z_b)}\right)  \label{eq:A}\;,\\
	\Phi(r_b,z_b)	&= \frac{k r_b^2}{2 R(z_b)} - \eta (z_b) + k z_b \label{eq:Phi-als-F_von_R_eta_s}\;,
\end{align}
and the variable definitions listed in table~\ref{tab:Variables}.
The beam amplitude $A(r_b,z_b)$ is normalized such that an integration over the entire plane of incidence of the beam irradiance $I$ yields the beam power $P$:
\begin{equation}
	P= \int_{0}^\infty dr_b \, 2 \pi r_b   I := \frac{1}{2 Z} \int_{0}^\infty dr_b\, 2 \pi  r_b   |A(r_b,z_b)|^2\;.
\end{equation}
\begin{table}[tb]
\centering
\begin{tabularx}{\textwidth}{| l  X  p{3.3cm} | }
	\hline\hline
	parameter		& description & characterizing eq. \\
	\hline  \hline
	$k$			& wave number common for both beams	& $k= 2 \pi/\lambda$ \\
	$\lambda$	& wavelength &\\
	$\omega_{m,r}$ & angular frequency of meas. / ref. beam, used solely to define the heterodyne frequency $\Delta \omega$ & $\omega=c k = 2 \pi f $\\
	$\Delta \omega$ & angular heterodyne frequency & $\Delta \omega=| \omega_m-\omega_r |$ \\
	$z_{0,m,r}$	& Rayleigh range of meas. / ref. beam & $z_0= \pi \,w_0^2/\lambda$\\ 
	$w_{0,m,r}$	& waist size of meas. / ref. beam &$w_0= \sqrt{z_0\, \lambda/\pi}$ \\
	$w_{m,r}$	& laser spot size on detector & $ w=w_0 \sqrt{1+ (z/z_0)^2}$\\
	$R_{m,r}$ 	& radius of curvature of meas. / ref. beam & $R=z (1+(z_0/z)^2)$\\
	$q_{m,r}$		& q-parameter of meas. / ref. beam & $q=z+i z_0,\newline 1/q= 1/R - i \lambda/(\pi w^2)$\\ 
	$\eta_{m,r}$	& Gouy phase of meas. / ref. beam & $\eta = \arctan(z/z_0)$\\
	$P_{m,r}$		& power of meas. / ref. beam &\\
	$\bar P$		& (time averaged) power in the interferometer & $P_m+P_r$ \\
	$Z$			& impedance of the medium  &\\
	$z_{m,r}$		& distance from waist in direction of propagation for meas. / ref. beam &\\
	$r_{m,r}$		& cylindrical coordinate  of meas. / ref. beam& $r=\sqrt{x^2+y^2}$\\
	$(x, y, 0)^T$		& point on the detector surface &\\
	$(x_i, y_i, 0)^T_{m,r}$ & on the detector: incident point of  meas.  / ref. beam  &\\
	$(x_p, y_p z_p)^T_{m,r}$ & pivot point of meas.  / ref. beam rotation & \\
	$(n_x,n_y,n_z)^T_{m,r}$ & rotation axis of meas.  / ref. beam rotation &\\ 
	$\alpha_{m,r}$ 	& tilt angle of meas.  / ref. beam & \\
	\hline  \hline
\end{tabularx}
\caption{List of physical parameters.}
\label{tab:Variables}
\end{table}
The beat note in a heterodyne interferometer, or generally the power sensed by a detector with surface $S$ in a two beam laser interferometer is given by
\begin{align}
	P_S  &= \int \text{d}S |E_m +E_r|^2 \\
	  &= \int \text{d}S \frac{1}{2 Z}\, |A_r\,\exp(i \omega_r t - i \Phi_r)+A_m\,\exp(i \omega_m t - i \Phi_m)|^2\\
	  &= \int \text{d}S  \frac{1}{2 Z}\, (A_r^2+A_m^2) \left[1+ \frac{2 A_m A_r}{A_m^2+A_r^2}\cos(\Delta \omega t - \Delta \Phi) \right] \\
	  &= P_m + P_r \left[1+ \frac{1}{P_m+P_r}  \int \text{d}S  \frac{1}{2 Z}( 2 A_m A_r)\cos(\Delta \omega t - \Delta \Phi) \right]\\
	  &=: \bar P \left[1+  \frac{1}{\bar P}  \int \text{d}S 
		\frac{2 \sqrt{P_m P_r}}{\pi w_m w_r}\exp\left(-\frac{r_m^2}{w_m^2} -\frac{r_r^2}{w_r^2}\right) 
				2 \cos(\Delta \omega t - \Delta \Phi) \right]\\
	  &= \bar P \left[1+  \frac{2 \sqrt{ P_m P_r}}{\bar P}
		\int \text{d}S    \frac{2 }{\pi w_m w_r} \exp\left(-\frac{r_m^2}{w_m^2} -\frac{r_r^2}{w_r^2}\right) 
				 \cos(\Delta \omega t - \Delta \Phi) \right] \label{eq:P_eq_ExpCos_wo_cootrafo}\\				
	  &=: \bar P \left[1+A_P  \int \text{d}S\, I_{Ov} \right] \label{eq:OverlapInt1} \;,		 
\end{align}
where it was assumed that both beams have identical wave numbers: $k_m = k_r = k$. For a heterodyne interferometer, it is thus assumed, that the heterodyne frequency $\Delta \omega$ is small compared to both angular frequencies:  $\Delta \omega \ll \omega_{m,r}$. As shown in \cite{Wanner2012}, the detected power $P_S$ can be generally expressed in the form
\begin{equation}
	P_S  = \bar P \left[1+c \cos(\Delta \omega t + \phi) \right]\;, \label{eq:BeatNote-Standard}
\end{equation}
such that for any given surface $S$ a specific contrast $c$ and phase $\phi$ are sensed by the detector. We described methods to compute these parameters numerically for arbitrary interferometers in \cite{Wanner2012}. It is possible to compute these signals analytically for infinitely large single element photodiodes, as we will show below.
The equations shown so far describe two coaligned beams impinging on  the detector in normal incidence. In order to allow each beam to be shifted and tilted arbitrarily, a coordinate transformation for each beam needs to be performed. For this transformation, let the detector plane be located at z=0. Let the point $(x_i, y_i, 0)^T_{m,r}$ be the initial beam displacement, which is intersection point of the beam axis with the detector plane before a rotation is applied ($\alpha_{m,r}=0$). Assume then, that each beam is rotated around a remote pivot $(x_p, y_p, z_p)^T_{m,r}$ and an arbitrary axis $\hat e_n=(n_x,n_y,n_z)^T_{m,r}$ with $\hat e_n.\hat e_n^T=1$. The coordinate transformation is then:
\begin{equation}
\begin{pmatrix}
		x \\ 	y\\ 	z
	\end{pmatrix}_{m,r}
	=
	M_{m,r} . \left[
	\begin{pmatrix}
		x \\ 	y\\ 	0
	\end{pmatrix}	-		
	\begin{pmatrix}
		x_p\\ 	y_p\\ z_p
	\end{pmatrix}_{m,r}
	\right]
	+		
	\begin{pmatrix}
		x_p\\ 	y_p\\ z_p
	\end{pmatrix}_{m,r}
	 -
	\begin{pmatrix}
		x_i\\ 	y_i\\ 	0
	\end{pmatrix}_{m,r}
	\label{eq:CooTrafo}
\end{equation}
where $M$ is the rotation matrix around an arbitrary axis $(n_x,n_y,n_z)^T_{m,r}$
\begin{align}
	&M_{m,r}:= \\
& \small	\begin{pmatrix}
	n_x^2 (1 - \cos(\alpha)) + \cos(\alpha) & n_x n_y (1 - \cos(\alpha)) - n_z \sin(\alpha)&    n_x n_z (1 - \cos(\alpha)) + n_y \sin(\alpha)\\ 
    	n_y n_x (1 - \cos(\alpha)) + n_z \sin(\alpha) & n_y^2 (1 - \cos(\alpha)) + \cos(\alpha)&     n_y n_z (1 - \cos(\alpha)) - n_x \sin(\alpha) \\
    	n_z n_x (1 - \cos(\alpha)) - n_y \sin(\alpha) &     n_z n_y (1 - \cos(\alpha)) + n_x \sin(\alpha) & n_z^2 (1 - \cos(\alpha)) + \cos(\alpha) \\
	\end{pmatrix}_{m,r}\hspace{-2mm}. \nonumber
\end{align}
Applying this coordinate transformation means that any $r_{m,r}, z_{m,r}$ in Eq.~(\ref{eq:P_eq_ExpCos_wo_cootrafo}) needs to be substituted according to Eq.~(\ref{eq:CooTrafo}). Due to their definition (see Tab.~\ref{tab:Variables}), also the spot sizes on the detector $w_{m,r}$, radii of curvature $R_{m,r}$ and Gouy phases $\eta_{m,r}$ are subject to the transformation. However, it is usually sufficient to assume that these parameters are constant on the detector surface. We therefore do not perform the coordinate transformation on $w_{m,r}, R_{m,r}, \eta_{m,r}$ and keep them as constants. By some lengthy algebraic manipulations which can be carried out by a standard analytical software tool, the overlap integrand can then be brought to the form
\begin{align}
	I_{Ov} &= A_0 \exp[-(A_1 x^2 + A_2 x y + A_3 x + A_4 y^2 + A_5 y + A_6)] \cos[  B_1 x^2 + B_2 x y+ B_3 x  + B_4 y^2 + B_5 y + B_6] \nonumber\\
	&=A_0 \exp\left[-
	\begin{pmatrix}
		x\\ 	y\\ 1\\
	\end{pmatrix}^T
	\begin{pmatrix}
		A_1 &  A_2 & A_3 \\ 
		0 & A_4 & A_5 \\ 
		0 & 0 & A_6 \\
	\end{pmatrix}
	\begin{pmatrix}
		x\\ 	y\\ 1\\
	\end{pmatrix}
	\right]
	\cos\left[
		\begin{pmatrix}
		x\\ 	y\\ 1\\
	\end{pmatrix}^T
	\begin{pmatrix}
		B_1 &  B_2 & B_3 \\ 
		0 & B_4 & B_5 \\ 
		0 & 0 & B_6 \\
	\end{pmatrix}
	\begin{pmatrix}
		x\\ 	y\\ 1\\
	\end{pmatrix}
	\right]\;.	\label{eq:I_Ov}
\end{align}
An explicit definition of the coefficients $A_n, B_n$ is listed in Tab.~\ref{tab:Coefficients} for the example that the reference beam impinges normally ($\alpha_r=0$), and the measurement beam rotates around the y-axis ($(n_x,n_y,n_z)=(0,1,0), \alpha_m=\alpha$). 
It is now possible to solve the overlap integral in Eq.~(\ref{eq:OverlapInt1}) and extract the interferometric contrast $c$ and phase $\phi$ for this general case.
\begin{table}[tb]
\centering
\begin{tabularx}{\textwidth}{| p{1.1cm}  X  | }
	\hline\hline
	variable		& substituted formula \\
	\hline  \hline
	$A_P$	& $2 \sqrt{P_m P_r}/(P_m + P_r)$ \\
	$A_0$	& $2/(\pi w_m w_r)$\\
	$A_1$	& $1/w_r^2 +\cos(\alpha_m)^2/w_m^2 $\\
	$A_2$	& 0 \\
	$A_3$	& $-2 x_{i,r}/w_r^2 - 2 x_{i,m} \cos(\alpha_m)^2/w_m^2$ \\
	$A_4$	& $ 1/w_m^2 + 1/w_r^2$ \\
	$A_5$	& $ -(2 y_{i,m})/w_m^2 -(2 y_{i,r})/w_r^2$ \\
	$A_6$ 	& $ (x_{i,m}^2 \cos(\alpha_m)^2+y_{i,m}^2)/w_m^2 +(x_{i,r}^2+ y_{i,r}^2)/w_r^2 $\\
	$B_1$	& $ -k/(2 R_r) + k \cos(\alpha_m)^2/(2 R_m) $\\
	$B_2$	& 0\\
	$B_3$	& $k x_{i,r}/R_r - k x_{i,m} \cos(\alpha_m)^2/R_m - k \sin(\alpha_m)$ \\
	$B_4$	& $ k/(2 R_m) - k/(2 R_r)$ \\
	$B_5$	& $ -k y_{i,m}/R_m + k y_{i,r}/R_r$ \\
	$B_6$	& $ -k(x_{i,r}^2+y_{i,r}^2)/(2 R_r)  + k( y_{i,m}^2+x_{i,m}^2 \cos(\alpha_m)^2)/(2 R_m)  - k( z_m- z_r) + \Delta\omega t - \eta_m + \eta_r + + k x_{i,m} \sin(\alpha_m) $\\
	\hline  \hline
\end{tabularx}
\caption{List of substitutions in the overlap integrand $I_{Ov}$ (Eq.~(\ref{eq:I_Ov})) for the special case $(n_x, n_y, n_z)_m=(0,1,0)$ and $\alpha_r=0$. Substitutions for the general case are given by substituting $r_m$ and $r_r$ in Eq.~(\ref{eq:P_eq_ExpCos_wo_cootrafo}) according to Eq.~(\ref{eq:CooTrafo}) and equating the coefficients in the overlap integrand Eq.~(\ref{eq:I_Ov}).}
\label{tab:Coefficients}
\end{table}
%
\section{Solving the overlap integral and extracting phase and contrast} \label{sec:solve_OvI}
In order to reduce the computational complexity, the overlap integrand $I_{Ov}$ can be extended to the complex domain:
\begin{align}
	I_{Ov} 
	&=:\, \Re \left(I_{Ov}^c \right) \\
	I_{Ov}^c &=A_0 \exp\left[-
	\begin{pmatrix}
		x\\ 	y\\ 1\\
	\end{pmatrix}^T
	\begin{pmatrix}
		A_1 &  A_2 & A_3 \\ 
		0 & A_4 & A_5 \\ 
		0 & 0 & A_6 \\
	\end{pmatrix}
	\begin{pmatrix}
		x\\ 	y\\ 1\\
	\end{pmatrix}
	\right]
	\exp\left[i 
		\begin{pmatrix}
		x\\ 	y\\ 1\\
	\end{pmatrix}^T
	\begin{pmatrix}
		B_1 &  B_2 & B_3 \\ 
		0 & B_4 & B_5 \\ 
		0 & 0 & B_6 \\
	\end{pmatrix}
	\begin{pmatrix}
		x\\ 	y\\ 1\\
	\end{pmatrix}
	\right]\\ 
	&=A_0 \exp\left[-
		\begin{pmatrix}
		x\\ 	y\\ 1\\
	\end{pmatrix}^T
	\begin{pmatrix}
		C_1 &  C_2 & C_3 \\ 
		0 & C_4 & C_5 \\ 
		0 & 0 & C_6 \\
	\end{pmatrix}
	\begin{pmatrix}
		x\\ 	y\\ 1\\
	\end{pmatrix}
	\right]\;.	
\end{align}
with $C_l= A_l - i B_l$ for $l=1,2,...,6$. 
The complex overlap integral $O_v^c$ can then be computed analytically for an infinite detector (S is the xy-plane):
\begin{align}
	O_v^c &:=  \int \text{d}S\, I_{Ov}^c \label{eq:def_Ovc}\\
	&= \frac{2 \pi \exp\left[-C_6 +
		\frac{ C_1 C_5^2+C_3^2 C_4 -C_2 C_3 C_5 }
	{4 C_1 C_4-C_2^2 } \right] }
	{ \sqrt{4 C_1 C_4 - C_2^2}}
\end{align}
provided that 
\begin{equation}
	\Re\left[4 C_4 - \frac{C_2^2}{C_1} \right]>0\;.
\end{equation}
The phase can be extracted in various ways, all of which result in the same expression. One option is to neglect the time oscillation ($t=0$) and then derive the phase as the argument of $O_v^c$ (this can e.g. be seen by combining equation (\ref{eq:OverlapInt1}), (\ref{eq:BeatNote-Standard}), (\ref{eq:I_Ov}), and (\ref{eq:def_Ovc})):
\begin{align}
	\phi 	&= \arg(O_v^c\big|_{t=0})= \arctan\left(\Im(O_v^c\big|_{t=0}),\Re(O_v^c\big|_{t=0})\right) \\
		&= \arctan(\tan( D_1)) = \text{mod}(D_1,\pi)-\pi/2
	 \label{eq:Result-phi}	
\end{align}
and $D_1$ is defined by:
\begin{equation}
	D_1:= B_6 +\frac{1}{2} \arg(D_2) + \frac{\Im(D_2 D_3^*)}{|D_2|^2}\label{eq:D_1}
\end{equation}
with
\begin{align}
	D_2 &:= C_2^2 - 4 C_1 C_4\\
	D_3 &:=C_3^2 C_4 - C_2 C_3 C_5 + C_1 C_5^2\;.
\end{align}
The contrast can be computed for instance by combining Eq.~(\ref{eq:OverlapInt1}) and Eq.~(\ref{eq:BeatNote-Standard}) and using the phase from Eq.~(\ref{eq:Result-phi}):
\begin{align}
	c 	&= \frac{A_P \Re(O_v^c)}{\cos(\Delta \omega t + \phi)}\\
		&=	2 \pi A_0 A_P \frac{\exp \left[-A_6 -\frac{\Re(D_2 D_3^*)}{|D_2|^2}\right]}{\sqrt{|D_2|}}\;.  \label{eq:Result-C-general}
\end{align}
%
\section{Comparison with numerical values}\label{sec:Verification}
\begin{figure}[tb]
	\includegraphics*[trim=22mm 20mm 14mm 24mm, width=0.5 \textwidth]{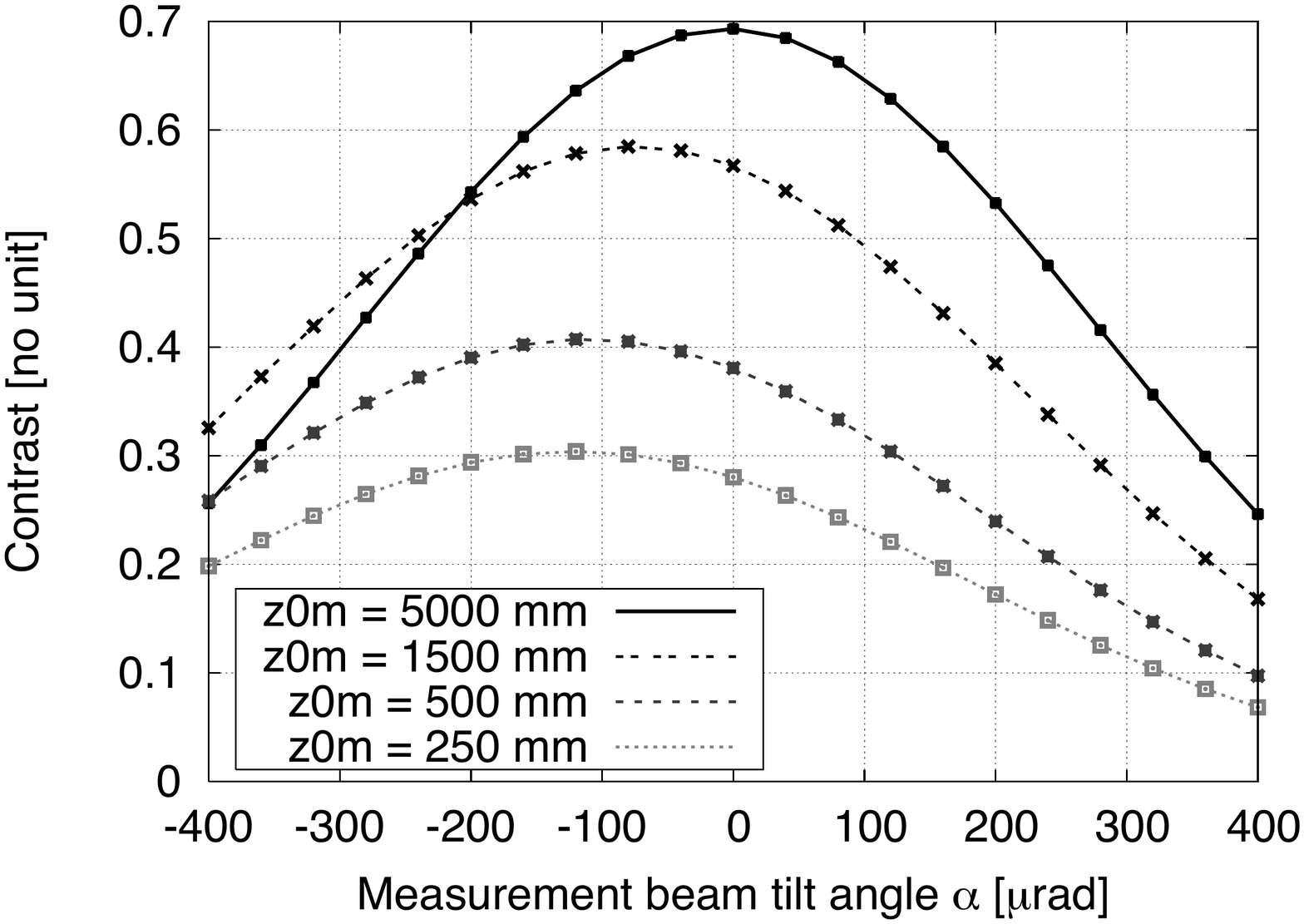}	
	\includegraphics*[trim=22mm 20mm 14mm 24mm, width=0.5 \textwidth]{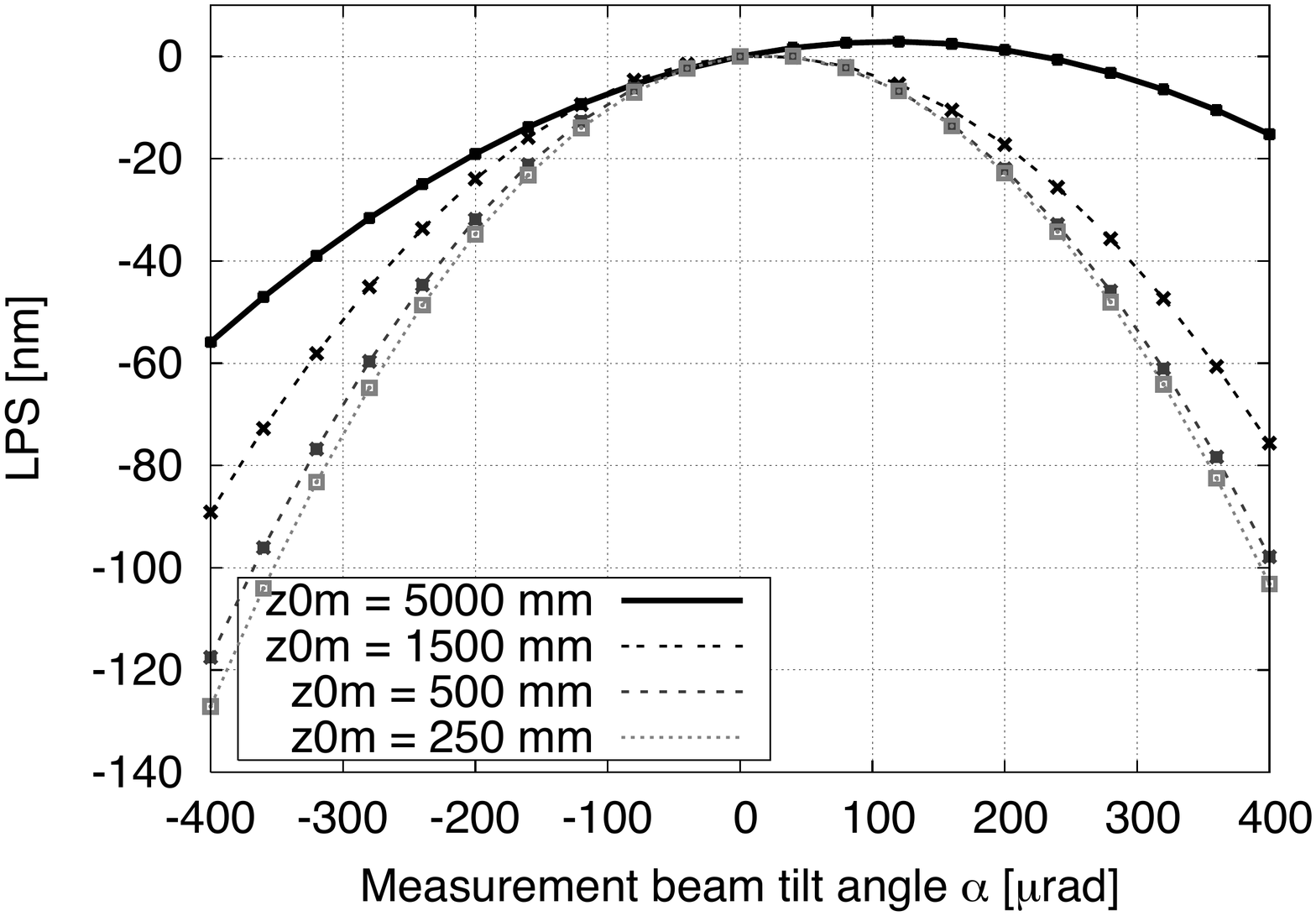}
	\caption{Resulting fringe visibility (contrast, left) and differential phase converted to length (longitudinal path length signal  $\text{LPS}= (\phi -\phi|_{\alpha_m=0})/k$, right). The solid lines were generated with Eq.~(\ref{eq:Result-C-general}) and Eq.~(\ref{eq:Result-phi}) respectively, the matching dots show numerical results generated with IfoCAD. }
	\label{fig:Verification}
\end{figure}

In order to verify the equations given above, we computed phase and contrast analytically using Eq.~(\ref{eq:Result-C-general}) and Eq.~(\ref{eq:Result-phi}) and compared them with numerical results from IfoCAD \cite{Kochkina2012} for a set of arbitrary values (listed below). The results match very well, as shown in fig~\ref{fig:Verification}. Here, the right hand side graph shows the interferometric phase converted to a length: the longitudinal pathlength signal 
\begin{equation}
\text{LPS}:= \frac{1}{k} (\phi -\phi|_{\alpha_m=0})\;. \label{eq:LPS}
\end{equation}
The graphs were generated with the following assumptions: 
$ k = 2 \pi/(1064\,\text{nm})$, 
$P_m=0.3\,$mW, 
$P_r=0.7\,$mW, 
$x_{i,m} = -400\,\upmu\text{m}$, 
$y_{i,m} = 300\,\upmu\text{m}$, 
$x_{i,r}=250\,\upmu\text{m}$, 
$y_{i,r} = -100\,\upmu\text{m}$, 
$z_r = 0.03452\,\text{m}$, 
$z_m = 3.768\,\text{m}$, 
$z_{0,r} = 3.137\,\text{m}$, 
rotation axis $(n_x,n_y,n_z)_m=(0,1,0), (n_x, n_y, n_z)_r=(0.5547,0.83205,0)$, 
pivot points $(p_x,p_y,p_z)_m=(12\,\upmu\text{m},50\,\upmu\text{m},4\,\text{mm}), (p_x,p_y,p_z)_r=(-7\,\upmu\text{m},200\,\upmu\text{m},-2\,\text{mm})$ 
and a reference beam angle $\alpha_r = 50\,\upmu$rad. The measurement beam angle was varied and contrast and LPS computed for different Rayleigh ranges of the measurement beam: $z_{0,m} =$ (5000\,mm, 1500\,mm, 500\,mm, 250\,mm). 
%
\section{Some useful special cases}\label{sec:Examples}
If both beams impinge with zero angle in the center of the detector ($\alpha_{m,r} = x_{i,m} = x_{i,r}=y_{i,m}=y_{i,r} = 0$), the contrast given in Eq.~(\ref{eq:Result-C-general}) takes the following form:
\begin{align}
	c	&= A_P \frac{2 \sqrt{z_{0,m} z_{0,r}}}{\sqrt{(z_{0,m} + z_{0,r})^2 + (z_m - z_r)^2}} \\
		&= A_P \frac{2}{w_m w_r \sqrt{\left(\frac{k}{2 R_m} -\frac{k}{2 R_r}\right)^2 + \left(\frac{1}{w_m^2} + \frac{1}{w_r^2}\right)^2}}
\end{align}
which was also found in \cite[Eq. (5.1)]{Dahl2013}. In the case of matched beam parameter ($z_m=z_r=z, z_{0,m}=z_{0,r}=z_0$ or equivalently $w_m = w_r=w, R_m=R_r=R$) normal incidence of both beams ($ \alpha_{m,r} =0$) we find:
\begin{align}
	c	&= A_P \exp\left[-k\frac{ (x_{i,m} - x_{i,r})^2 + (y_{i,m} - y_{i,r})^2}{4 z_0}\right]\\
		&= A_P \exp\left[-(4 R^2 + k^2 w^4) \frac{ (x_{i,m} - x_{i,r})^2 + (y_{i,m} - y_{i,r})^2}{8 R^2 w^2}\right]\;.
\end{align}

For us, a typical problem is the effect of beam tilt on the interferometric phase readout: $\phi(\alpha)$. We assume here a well aligned static reference beam ($\alpha_r =  x_{i,r} = y_{i,r} = 0$). The measurement beam is initially aligned (i.e. $x_{i,m}=y_{i,m}=0$), and then rotates around an arbitrary pivot point where the y-axis is chosen as rotation axis $(n_x,n_y,n_z)_m=(0,1,0)$. Furthermore, nearly equal beam parameters are assumed ($z_{0,r}=z_0, z_{0,m}=z_0 + \Delta z_0$ and $ z_{r}=z, z_{m}=z + \Delta z$ or $w_r=w, w_m=w+\Delta w, R_r =R, R_m = R+\Delta R$). 
The resulting longitudinal pathlength readout signal (Eq.~(\ref{eq:LPS})) expanded up to second order in the measurement beam angle $\alpha$ and first order in either variation $\Delta$ is:
\begin{equation}
	\text{LPS} (\alpha, \Delta z_0, \Delta z) \approx   \frac{\alpha^2}{k} \left( \frac{z}{4 z_0} + \Delta z_0 \frac{2 k z_0 (z_p + z)  - z}{8 z_0^2} +\Delta z \frac{k ((z+  z_p)^2- z_0^2) + z_0  }{8 z_0^2} \right) \label{eq:LPS_a_Dz0_Dz}
\end{equation}
and to first order differences of the radius of curvature $\Delta R$ and spot size $\Delta w$ (with $w_r=w, w_m=w_r+\Delta w, R_r =R, R_m = R+\Delta R$):
\begin{equation}
	\text{LPS} (\alpha, \Delta w, \Delta R)  \approx 
	\alpha^2  \left(   \frac{w^2}{ 8 R } +  \Delta R \frac{-2 R^2 (2 z_p^2 + w^2) +   k^2 w^4 (z_p + R)^2 } {32 R^4} + \Delta w \frac{w^2 + 4 z_p (z_p + R)}{8 R w}\right)\;.
\end{equation}

\section{Summary and Conclusions}
We have derived analytical equations for the phase and contrast of two arbitrary interfering Gaussian beams. We showed for a very general example perfect agreement with numerical results from IfoCAD \cite{Kochkina2012}. We showed reduced equations for special cases and compared to one known special case results.\\

The equations given here can be used to predict phase changes in interferometers and the contrast (fringe visibility), provided that a large single element detector is used. If the detector is not large compared to three beam radii and clipping of the Gaussian beams is expected, these equations should be handled with care. The reader should also be aware, that beam clipping for instance on the insensitive slit of a quadrant detector might change the detected phase considerably. For cases where beam clipping might occur, the well known numerical methods need to be used instead.

\section*{Acknowledgments}
We thank Marie-Sophie Hartig, S\"onke Schuster, and Michael Tr\"obs for valuable discussions. We gratefully acknowledge support by Deutsches Zentrum f\"{u}r Luft- und Raumfahrt (DLR) with funding of the Bundesministerium f\"{u}r Wirtschaft und Technologie with a decision of the Deutschen Bundestag (DLR project reference No 50 OQ 1301) and thank the Deutsche Forschungsgemeinschaft (DFG) for funding the Cluster of Excellence QUEST -- Centre for Quantum Engineering and Spacetime Research. 

\end{document}